\documentclass[acmsmall]{acmart}

\usepackage{algorithmic}
\usepackage{graphicx}
\usepackage{textcomp}
\usepackage{xcolor}
\usepackage{comment}
\usepackage{hyperref}
\usepackage{bm}
\usepackage{enumitem}
\usepackage{appendix}

\usepackage{amsmath}
\usepackage{lipsum}
\usepackage{multirow}
\usepackage[justification=centering]{caption}
\usepackage{caption}
\usepackage{url}
\usepackage{listings}
\usepackage{subfigure}
\usepackage{array}
\usepackage{pifont}
\usepackage{fontawesome5} 
\usepackage{colortbl}

\usepackage[most]{tcolorbox}
\usepackage{listings}
\usepackage{xcolor}
\usepackage{caption}

\lstdefinestyle{javastyle}{
  language=Java,
  basicstyle=\ttfamily\footnotesize,
  keywordstyle=\bfseries,
  commentstyle=\itshape\color{gray!70},
  stringstyle=\color{gray!50!black},
  showstringspaces=false,
  tabsize=2,
  columns=fullflexible,
  keepspaces=true,
  frame=single,
  framesep=3pt,
  framerule=0.4pt,
  breaklines=true
}

\newcommand{\badgeIR}[1]{\textsf{\colorbox{blue!10}{\footnotesize\textbf{IR}}}}
\newcommand{\badgeUR}[1]{\textsf{\colorbox{green!12}{\footnotesize\textbf{UR}}}}

\tcbset{
  colframe=black!10,
  colback=white,
  coltitle=black,
  fonttitle=\bfseries,
  boxrule=0.4pt,
  left=6pt,right=6pt,top=6pt,bottom=6pt,
  title style={left color=black!3,right color=black!1}
}

\AtBeginDocument{%
  }
\newtcolorbox{boxK}{
    top=2pt,
    bottom=2pt,
    left=2pt,
    right=2pt,
    boxrule = 0pt,
    toprule = 0pt, 
    colback=gray!15,   
    colframe=white     
}

\setcopyright{acmlicensed}
\copyrightyear{2018}
\acmYear{2018}
\acmDOI{XXXXXXX.XXXXXXX}

\acmJournal{PACMSE}
\acmVolume{37}
\acmNumber{4}
\acmArticle{111}
\acmMonth{8}

\def\eg{\textit{e.g.,} }
\def\ie{\textit{i.e.,} }

\newcommand{\ours}[1]{\textsc{UserTrace}}

\begin{document}

\author{Dongming Jin}
\email{dmjin@stu.pku.edu.cn}
\affiliation{%
  \institution{Peking University}
  \city{Beijing}
  \country{China}
}

\author{Zhi Jin}
\email{zhijin@pku.edu.cn}
\authornote{Zhi Jin is the Corresponding author}
\affiliation{
  \institution{Peking University and Wuhan University}
  \city{Beijing}
  \country{China}
}

\author{Yiran Zhang}
\email{yiran002@e.ntu.edu.sg}
\affiliation{
  \institution{Nanyang Technological University}
  \city{Singapore}
  \country{Singapore}
}

\author{Zheng Fang}
\affiliation{%
  \institution{Peking University}
  \city{Beijing}
  \country{China}
}

\author{Linyu Li}
\affiliation{%
  \institution{Peking University}
  \city{Beijing}
  \country{China}
}

\author{Yuanpeng He}
\affiliation{%
  \institution{Peking University}
  \city{Beijing}
  \country{China}
}

\author{Xiaohong Chen}
\affiliation{
  \institution{East China Normal University}
  \city{Shanghai}
  \country{China}
}

\author{Weisong Sun}
\affiliation{
  \institution{Nanyang Technological University}
  \city{Singapore}
  \country{Singapore}
}

\title{\ours{}: \underline{User}-Level Requirements Generation and \underline{Trace}ability Recovery from Software Project Repositories}

\begin{abstract}
Software maintainability critically depends on high-quality requirements descriptions and explicit traceability between requirements and code. Although automated code summarization (ACS) and requirements traceability (RT) techniques have been widely studied, existing ACS methods mainly generate implementation-level (\ie developer-oriented) requirements (IRs) for fine-grained units (\eg methods), while RT techniques often overlook the impact of project evolution. As a result, user-level (\ie end user-oriented) requirements (URs) and live trace links remain underexplored, despite their importance for supporting user understanding and for validating whether AI-generated software aligns with user intent. 

To address this gap, we propose \ours{}, a multi-agent system that automatically generates URs and recovers live trace links (from URs to IRs to code) from software repositories. \ours{} coordinates four specialized agents (\ie Code Reviewer, Searcher, Writer, and Verifier) through a three-phase process: structuring repository dependencies, deriving IRs for code units, and synthesizing URs with domain-specific context. Our comparative evaluation shows that \ours{} produces URs with higher completeness, correctness, and helpfulness than an established baseline, and achieves superior precision in trace link recovery compared to five state-of-the-art RT approaches. A user study further demonstrates that \ours{} helps end users validate whether the AI-generated repositories align with their intent.

\end{abstract}



\maketitle

\section{Introduction}

In the era of AI-assisted software development, a multitude of products (\eg Claude Code~\cite{ClaudeCode} and Kiro~\cite{Kiro}) can generate an entire software project from only a simple user requirements description. In this context, the maintainability attributes of software become increasingly vital, as they determine the long-term sustainability, evolution, and alignment of the generated systems with user intents. The alarming statistics from TechRadar Pro's AI Speed Trap Report~\cite{TapReport} reveal that nearly 66\% of organizations face critical challenges in maintainability, and SlashData notes that 46\% of developers report missing or insufficient requirements documentation~\cite{SlashData}. Furthermore, reports by Verified Market underscore an expanding trend in the software maintenance service market and may reach 180 billion USD by 2033~\cite{VerifiedMarket}. Against such a backdrop, ensuring software maintainability has become a critical concern, commanding attention in both academia and industry~\cite{schnappinger2020defining,jin2022evaluating,jolak2025empirical}. \textbf{In practice, ensuring maintainability critically depends on producing high-quality requirements descriptions and establishing explicit traceability links between requirements and code.} 

Despite this, writing requirements descriptions and maintaining traceability in software projects remains manual and labor-intensive. To alleviate human effort, the software engineering community has proposed various automated code summarization (ACS)~\cite{steidl2013quality,zhang2022survey} and requirements traceability (RT) techniques~\cite{torkar2012requirements}. Specifically, ACS approaches employ deep learning techniques~\cite{wan2018improving,wang2020reinforcement} or recent LLMs~\cite{sun2024source,geng2024large,sun2023automatic} to generate textual descriptions for fine-grained code units (\eg functions or classes), while RT approaches establish links between requirements and code based on information retrieval~\cite{gao2022propagating,kuang2019using}, deep learning~\cite{guo2017semantically,lin2021traceability}, and recent LLMs~\cite{fuchss2025lissa}. However, existing ACS methods primarily produce implementation-level requirements (IRs), which are developer-oriented and describe what the code does rather than why it is needed. Meanwhile, RT methods often construct static trace links and overlook the impact of project evolution, which restricts their ability to preserve valid and meaningful trace links as systems evolve. These leave a critical gap in summarizing user-level (\eg end user-oriented) requirements (URs) and maintaining live trace links.

Beyond IRs, producing URs is crucial for supporting user-level understanding and ensuring that software projects remain aligned with user goals~\cite{spijkman2022back}, especially in the era of AI-assisted software development, where end users with limited programming expertise can develop software using AI products with simple user intents. The key differences between URs and IRs lie in their abstraction level, target audience, and descriptive focus. IRs describe fine-grained units from the perspective of implementation, whereas URs capture higher-level system objectives from the perspective of end users. For example, an IR may state that ``\textit{the function \texttt{checkPassword()} verifies a user’s password hash against the stored hash in the database}'' while the corresponding UR would state that ``\textit{the system shall allow users to securely log into their accounts by verifying their credentials}''. As this example illustrates, IRs emphasize how the system is implemented, while URs articulate what the system is expected to achieve. Besides, to fully support maintainability, it is also necessary to establish live trace links that connect URs to IRs and ultimately to code, thereby enabling consistent evolution and user-aligned validation of software systems.

\begin{figure}
    \centering
    \includegraphics[width=0.9\linewidth]{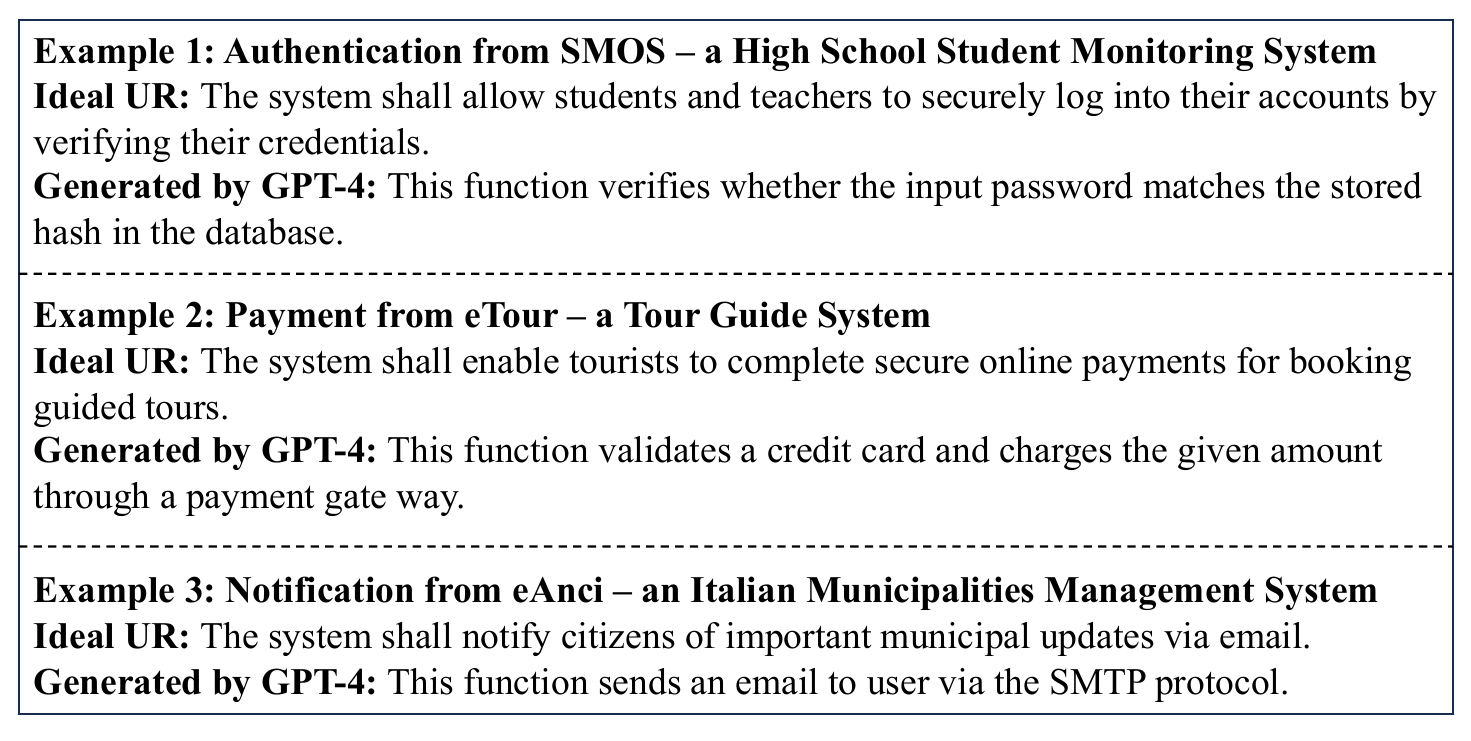}
    \caption{Three illustrative cases comparing GPT-4 generated requirements with ideal URs.}
    \label{fig:examples}
\end{figure}

Advanced Large Language Models (LLMs) like GPT-4~\cite{achiam2023gpt} have demonstrated remarkable capabilities in natural language requirements generation~\cite{jin2024mare,jin2024evaluation,jin2025iredev} and project repositories understanding~\cite{wang2025repomaster,yang2025docagent}. While they can produce contextually fluent requirements texts~\cite{lian2025incorporating} and recover basic traceability links across artifacts~\cite{fuchss2025lissa}, our research reveals a notable shortfall: in repository-level (\ie long context) scenarios and without domain-specific business knowledge, the LLMs tend to hallucinate non-existent code units and to generate descriptions that resemble IRs rather than URs. As demonstrated by three cases in Figure~\ref{fig:examples}, when tasked with code units from different software systems, GPT-4 regularly produces implementation-oriented descriptions. Although technically sound, these descriptions often fail to articulate the underlying purpose, focusing on how the code works rather than why the system exists, and include many implementation details that are difficult for end users to understand and validate. For example, in the first case study, GPT-4 describes the password verification process at the function level but overlooks the broader user-level requirement that the system must provide secure authentication for account access.

We identify three primary challenges that drive these shortcomings. \textbf{(1) Context Identification and Dependency Navigation.} Large software repositories contain many interdependent components, making it non-trivial for LLMs to accurately identify relevant context and capture relationships among code units. It requires strategic context and dependency management. \textbf{(2) Domain-specific Business Information Integration.} User-level requirements typically involve business rules, user workflows, or application specific semantics that are not explicitly encoded in source code. In practice, engineers often guarantee the user-level requirements by utilizing related business information. \textbf{(3) User Intent Abstraction.} Moving from code to IRs and further to URs requires abstraction from low-level implementation details to higher-level system objectives. Current LLMs struggle with this abstraction process, lacking sufficient generalization capability to capture user intent beyond technical descriptions.

Inspired by this insight, our design philosophy is to leverage a collaborative, role-specialized framework to overcome the identified challenges. To address context and dependency management, the methodology organizes repositories into structured graphs, ensuring that each generating step operates on the most relevant context. To handle domain knowledge integration, we incorporate retrieval mechanisms that enrich requirement generation with domain-specific business information beyond source code. To close the user intent abstraction gap, the methodology employs abstraction-oriented agents that transform low-level IRs into higher-level URs.


Building on this design philosophy, we propose \ours{}, a multi-agent collaborative system that automatically generates URs and recovers live trace links (from URs to IRs to code) from software repositories. \ours{} integrates four specialized agents (\ie Code Reviewer, Searcher, Writer, and Verifier). Each agent is responsible for a distinct aspect of the overall workflow. \ours{} operates in three phases: structuring repositories, deriving IRs for code units, and synthesizing URs with domain-specific context and iterative feedback. Initially, \ours{} applies static analysis to construct a dual-level dependency graph (\ie file-level and component-level), 
followed by a topological sort to provide a contextualized processing order for subsequent processing. Subsequently, the Code Reviewer produces IRs for each code component and file based on the constructed graph. Finally, the Writer applies the Leiden community detection algorithm to cluster related files and generates URs by abstracting from IRs and incorporating domain-specific business context retrieved by the Searcher. The verifier evaluates the generated URs along multiple dimensions (\ie completeness) and provides iterative feedback to the Writer and Searcher until no further issues are identified according to the predefined evaluation criteria.

In our experiments, we initially evaluated all URs generated by \ours{} for each evaluated system as a requirements document and assessed their quality from three key attributes (\ie completeness, correctness, and helpfulness) using an LLM-as-a-Judge framework~\cite{liu2023g}~\cite{yang2025docagent}. We then regarded the generated URs as a requirements set and manually evaluated them with traditional metrics (\ie precision, recall, and F1 score). We compared our \ours{} against the hierarchical summarization method~\cite{dhulshette2025hierarchical} with three recent LLMs on three evaluated systems (Section~\ref{subsec:evalsys}). The quantitative results demonstrate that \ours{} substantially outperforms the baseline in producing high-quality URs. For traceability recovery, we further assessed the quality of the trace links between URs and the implementation established by \ours{}. \ours{} achieved the highest accuracy rate, surpassing five state-of-the-art RT baselines. In addition, we conducted a user study to evaluate the practical utility of \ours{}. Results demonstrate that \ours{} enables end users to more effectively validate whether AI-generated repositories align with their intent, thereby confirming its value in bridging the gap between user goals and implementation.

The contributions of our work are threefold:
\begin{itemize}
    \item To the best of our knowledge, this is the first work to explore URs generation and URs-to-code traceability from software project repositories, highlighting their importance for software maintainability in the era of AI-assisted development. 
    \item We propose \ours{}, a multi-agent collaborative system that automatically generates URs and recovers live trace links from software repositories by coordinating four specialized agents across three phases. 
    \item We conduct extensive experiments and a user study. Both qualitative and quantitative analyses demonstrate the effectiveness and practicality of our \ours{}. \ours{} produces URs with high quality and achieves superior precision in trace link recovery.
\end{itemize}

\section{Background: User-level Requirements (URs) and Motivation Example}

\subsection{User-level Requirements}


Software requirements can be broadly defined as the desired effects or outcomes that end users expect a software system to produce~\cite{pohl1996requirements}. They should capture what stakeholders want to achieve through the system. 
For example, a requirements description may be expressed as ``\textit{The system shall allow users to log into their accounts securely}''. However, in current automated code generation scenarios, the requirements descriptions provided as input primarily emphasize program logic and developer-oriented details (\eg ``\textit{Write a function to compare two strings}''). Obviously, such descriptions differ from the general definition of software requirements in terms of abstraction level, descriptive focus, and target audience. In this paper, we categorize requirements descriptions into two layers, \ie implementation-level requirements (IRs) and user-level requirements (URs). 

Specifically, user-level requirements are high-level descriptions of what a system is expected to accomplish, emphasizing user goals, business semantics, and overall purpose. Formally, an UR can be expressed as $U = f(G, C, B)$, where $G$ denotes user goals, $C$ represents contextual constraints, and $B$ refers to domain-specific business rules. In contrast, implementation-level requirements are developer-oriented descriptions derived from program behavior, \ie $I = g(P, C)$, where $P$ represents program logic and $C$ denotes contextual constraints. For example, in an authentication module, a IR may be expressed as ``\textit{the function checkPassword() compares a user's input against that stored hash retrieved from the database}''. While accurate at the implementation level, this description reflects only how the system works. The corresponding UR would instead state as ``\textit{The system shall allow users to securely log into their accounts by verifying their credentials}''.

The importance of generating URs is particularly pronounced in the era of AI-assisted software development. With the advent of LLM-powered generative tools, entire software projects can be produced from simple natural language intent descriptions, even by non-developers with limited experience. In this context, validating whether the generated repositories faithfully realize user goals becomes both critical and challenging. Compared to IRs, generating URs from software project repositories is more suitable and valuable for end users to validate alignment with their intent and to support further updates. Specifically, URs provide three essential benefits. \textbf{(1) Validation:} URs serve as a benchmark to assess whether AI-generated repositories align with stakeholder expectations. \textbf{(2) Traceability:} URs anchor cross-level links by connecting high-level goals to IRs and code, thereby supporting consistent evolution of system change. \textbf{(3) Maintainability:} URs establish stable abstractions beyond implementation details, ensuring that systems remain sustainable and comprehensible over time. In this sense, URs are a practical foundation for building trustworthy, user-aligned software in an AI-assisted software development paradigm.

\begin{figure}
    \centering
    \includegraphics[width=0.8\linewidth]{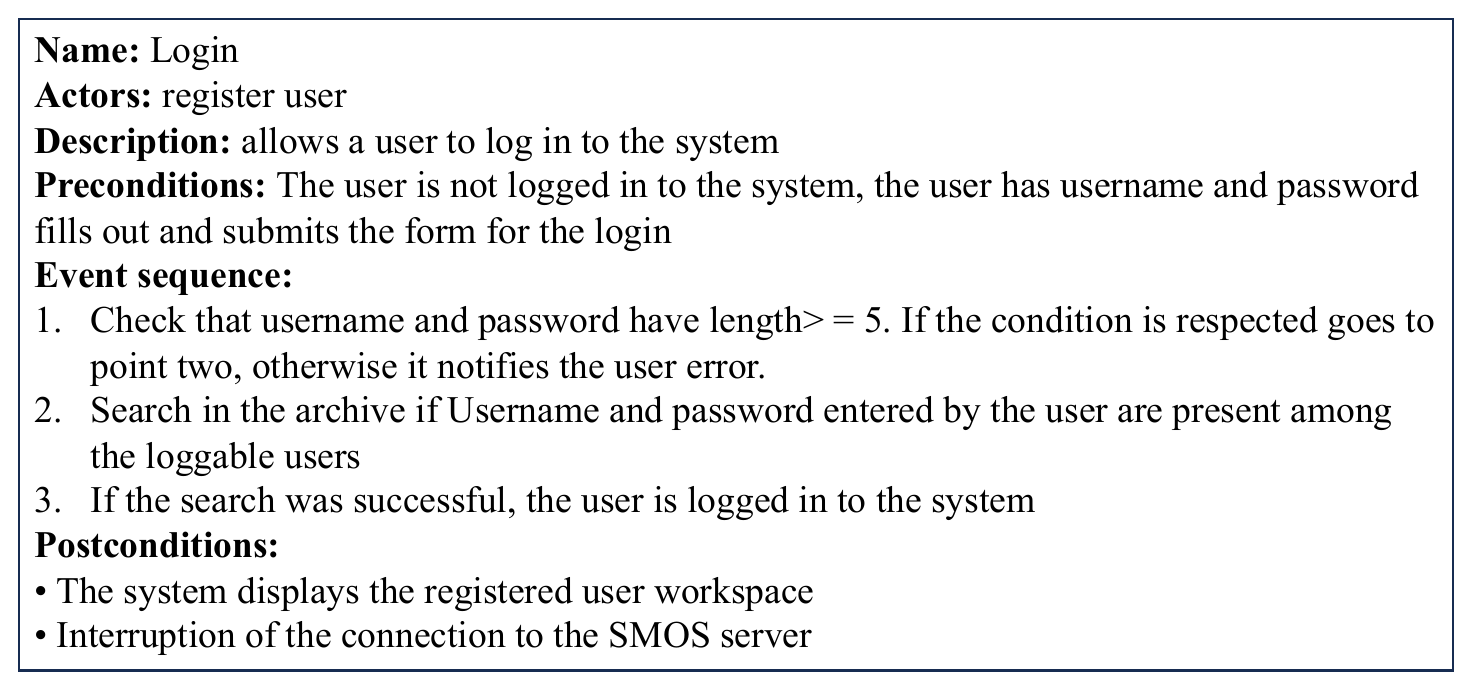}
    \caption{An Example of Use Case from the School Student Monitoring System (SMOS)}
    \label{fig:usecase}
\end{figure}

In practice, user-level requirements are often naturally expressed in the form of use cases~\cite{bittner2003use}, which describe system interactions from the perspective of end users. A use case typically specifies the actors involved, the goals they aim to achieve, and the scenarios through which these goals are realized. A use case contains six key attributes, \ie name, actors, description, preconditions, event flow, and exit conditions. Figure~\ref{fig:usecase} shows a use case in the SMOS system. Compared to IRs that focus on specific functions or code behavior, use cases capture the intent, context, and expected outcomes at a higher level of abstraction. They provide a concise yet powerful means for bridging communication between stakeholders and developers, ensuring that system functionalities remain aligned with business goals. In this paper, we adopt use cases as the representation of URs, serving both as the target format for automatic generation.


\subsection{Motivation Example}

To systematically generate and recover URs and trace links from software repositories, we propose leveraging IRs and dependency structures embedded in the repositories as a roadmap, gradually abstracting and transforming them into URs. Our intuition is that, while repositories primarily encode developer-oriented implementation logic, they also contain implicit signals of user intent. If properly extracted and abstracted, these signals can be transformed into URs.

\begin{figure}[t]
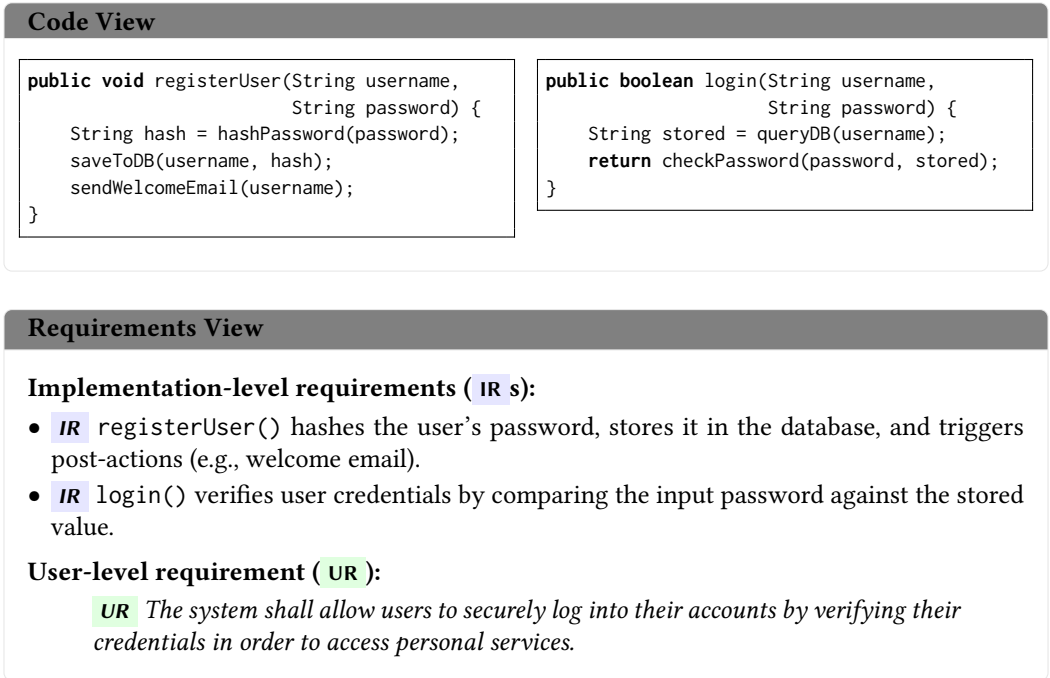

\centering

\begin{tcolorbox}[title=Code View, width=\linewidth]
  \begin{minipage}[t]{0.48\linewidth}
    \vspace{0pt}
\begin{lstlisting}[style=javastyle]
public void registerUser(String username, 
                         String password) {
    String hash = hashPassword(password);
    saveToDB(username, hash);
    sendWelcomeEmail(username);
}
\end{lstlisting}
  \end{minipage}\hfill
  \begin{minipage}[t]{0.48\linewidth}
    \vspace{0pt}
\begin{lstlisting}[style=javastyle]
public boolean login(String username, 
                     String password) {
    String stored = queryDB(username);
    return checkPassword(password, stored);
}
\end{lstlisting}
  \end{minipage}
\end{tcolorbox}

\vspace{0.2em}

\begin{tcolorbox}[title=Requirements View, width=\linewidth]
  \textbf{Implementation-level requirements (\badgeIR{}s):}
  \begin{itemize}[leftmargin=*, itemsep=0.15em]
    \item \textit{\badgeIR{}}~\texttt{registerUser()} hashes the user's password, stores it in the database, and triggers post-actions (e.g., welcome email).
    \item \textit{\badgeIR{}}~\texttt{login()} verifies user credentials by comparing the input password against the stored value.
  \end{itemize}

  \vspace{0.25em}
  \textbf{User-level requirement (\badgeUR{}):}
  \begin{quote}\itshape
    \badgeUR{}~The system shall allow users to securely log into their accounts by verifying their credentials in order to access personal services.
  \end{quote}
\end{tcolorbox}

\captionsetup{font=small}
\caption{An Example Demonstrating the Relationship Between Source Code, IRs, and URs}
\label{fig:motivation-example}
\end{figure}

Take the authentication module as an instance in Fig~\ref{fig:motivation-example}: IRs and URs are clearly related at first. The function \texttt{checkPassword()} implements password verification, \texttt{registerUser()} handles account creation, and \texttt{login()} manages session establishment. These IRs collectively describe how the system operates. However, taken together, they support a higher-level UR: ``The system shall allow users to securely log into their accounts by verifying their credentials''. This example highlights that recovering URs is not a simple synonym-matching or surface-level semantic similarity task. Instead, it requires deeper reasoning that spans code, context, and business knowledge. We distill four key insights that guide our design:


\begin{itemize}
    \item \textbf{Insight I (Context and Dependency Navigation):} 
    directly feeding an entire repository as context into LLMs often exceeds their context window limitations.
    
    \item \textbf{Insight II (Multiple Code Units Aggression):} A single UR often emerges only when the behaviors of multiple IRs are aggregated. For instance, the UR ``\textit{The system shall allow users to securely log into their accounts}'' cannot be derived from \texttt{checkPassword()} alone but requires integrating evidence from the registration and login modules.  
    \item \textbf{Insight III (Business Knowledge Integration):} URs may not be fully inferred from code. For example, ``\textit{The system shall allow users to securely access personal services}'' depends on external security policies, which must be incorporated into the abstraction process.  
    \item \textbf{Insight IV (User-Level Abstraction):} Moving from IRs to URs requires abstraction from low-level details to higher-level goals. While IRs describe the technical logic of password hashing and verification, the UR frames this as the broader objective of secure authentication for account access, aligning the system with end-user expectations.
\end{itemize}

These four insights illustrate that generating URs and establishing live traceability from repositories is a non-trivial task. It requires structured context management, cross-unit reasoning, domain knowledge integration, and abstraction mechanisms. These challenges directly motivate the design of \ours{}, a multi-agent system that collaborates to recover URs and their trace links.


\section{Approach}

In this section, we present \ours{}, a multi-agent collaborative system to generate URs and recover live URs-to-code trace links from software repositories. We formally defined the overview of our \ours{} and describe the details in the following sections, including three phases.

\subsection{Overview}
\begin{figure}
    \centering
    \includegraphics[width=0.99\linewidth]{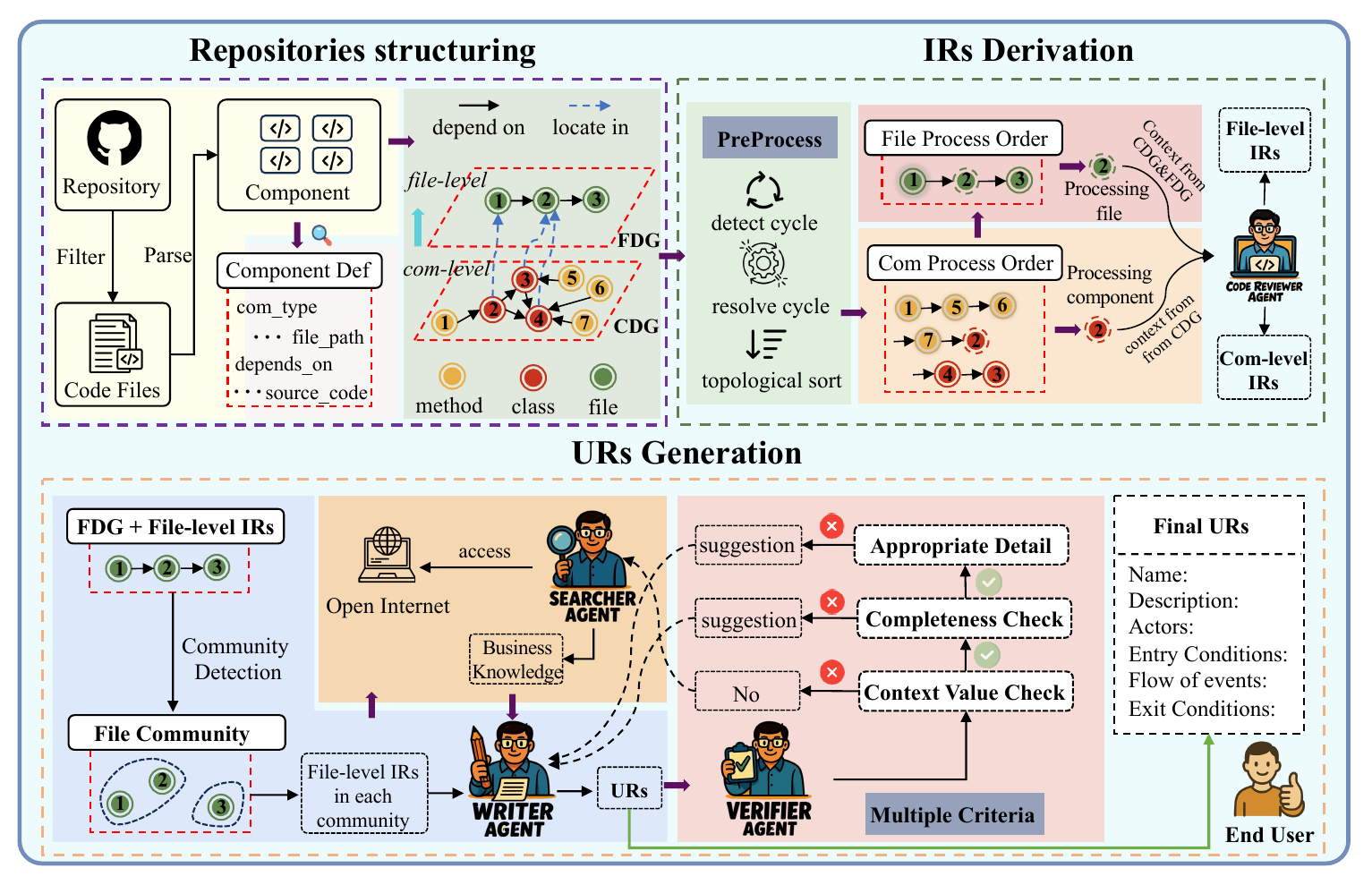}
    \caption{Overview of our \ours{}.}
    \label{fig:overview}
\end{figure}

\ours{} integrates four specialized agents (\ie Code Reviewer, Searcher, Writer, and Verifier). They collaborate across three phases: repository structuring, IR Derivation, and UR Generation. Figure~\ref{fig:overview} illustrates the overview of \ours{}.

\begin{itemize}
    \item \textbf{Repositories structuring:} parses components and their relationships from a repository to construct a dual-level dependency graph, including component-level and file-level. 
    \item \textbf{IR Derivation:} leverages the dual-level dependency graph to retrieve relevant context and prompt the Code Reviewer agent to drive IRs for each component and file. 
    \item \textbf{UR Generation:} prompts the Writer agent to abstract IRs into URs with the assistance of the Searcher agent (for retrieving domain business knowledge) and the Verifier agent(for iterative quality assessment and refinement).
\end{itemize}

\subsection{Repositories Structuring}
This phase aims to transform raw code repositories into structured representations that support subsequent requirement generation and trace recovery. Generating accurate and appropriately granular requirements presupposes a comprehensive understanding of the repository. However, directly feeding an entire repository as context into LLMs often exceeds their context window limitations, especially for large and structurally complex projects. To address this challenge, we abstract the repository into dual-level dependency graphs, including a component dependency graph (CDG) and a file dependency graph (FDG). This graph can provide an ordered processing structure for requirement generation and ensure that the most relevant dependencies can be retrieved as context when generating each requirement.

Specifically, \ours{} first filters out non-code files and performs static analysis on all source files. For each file, we parse its abstract syntax tree (AST) to identify core code components\footnote{Java projects cover class, interface, enum, method; Python projects cover class, function, method} and their dependencies. Each component is abstracted as a node with multiple key attributes, \eg component type, file path, depends on, and source code. Based on these nodes and their relations, \ours{} builds a directed graph (\ie the CDG), where an edge from node A to node B (A $\rightarrow$ B) indicates that A depends on B. Leveraging the mapping between components and files, \ours{} further construct the FDG at the file level, capture cross-file dependencies. 

This dual-layer dependency graph representation provides the foundation for subsequent phases. The CDG (or FDG) provides the component (or file) processing order and assists the related context retrieval for the IR derivation phase. The FDG also offers a high-level aggregation perspective for the UR generation phase, allowing \ours{} to synthesize the IRs into URs.



\subsection{IRs Derivation}
The goal of this phase is to derive IRs for all components and files in the given software repository, providing the semantic foundation for subsequent abstraction into URs. As a continuation of repository structuring, this phase emphasizes a dependency-aware and context-aware generation process. When processing a component or file, the IRs of its dependencies have already been produced and can thus serve as contextual input. 

The stage begins by preprocessing the CDG and FDG. Specifically, cycles are detected through strongly connected component analysis and resolved by selectively removing edges, which transforms the graphs into directed acyclic graphs (DAGs). Topological ordering is then applied to both graphs to ensure a valid generation order. This ordering guarantees that when deriving a IR for a component, all its direct dependencies have already been processed, allowing \ours{} to leverage one-hop dependency information rather than expanding entire software projects. This design avoids excessive contextual overhead and ensures the retrieved context remains focused and relevant. Based on this order, \ours{} employs the Code Reviewer agent to generate IRs at the component level. The Code Reviewer agent receives the component's source code together with the names and IRs of its dependent components, and produces a component-level IR. Once all components have been processed, \ours{} employs the Code Reviewer agent to aggregate all components within the same file and produce a file-level IR. 

Through this two-stage ``components-first, files-next'' derivation strategy, \ours{} can capture component requirements at the fine-grained level and form globally consistent abstraction at the file-level. This design ensures that IRs are compositional and coherent across different granularities, thereby providing a solid foudation for the generation of URs in the next phase.


\subsection{URs Generation}
This final phase of our \ours{} aims to generate URs by abstracting the previously derived file-level IRs, which bridge the gap between low-level implementation details and high-level user intent. Unlike IRs derivation, which primarily emphasizes dependency-aware reasoning within the repository, this URs generation stage emphasizes the integration of both repository-derived knowledge and external business knowledge. This process is further enhanced by a ``verify-then-feedback'' mechanism. This design ensures that the generated URs are meaningful from a business perspective and align with user goals.

Specifically, \ours{} employ three agents to perform this stage, \ie the Writer, the Searcher, and the Verifier. Firstly, the Writer agent receives the set of file-level IRs and applies a Leiden community detection algorithm on the FDG to partition files into semantically cohesive groups. Each community serves as a unit for UR generation, preventing context overload and preserving semantic boundaries. For each community, the writer synthesizes IRs into candidate URs that reflect broader functional goals. 
Before this process, the Searcher agent is invoked to identify missing contextual business knowledge in IRs and access the open internet engine to get this external knowledge (\ie relevant business concepts and domain-specific terminology). By combining IRs with this supplementary knowledge, the Writer agent produces URs. To ensure the quality and reliability of the generated URs, a Verifier agent is involved to evaluate the generated URs and give feedback to the Writer or Searcher agent. The verifier takes the generated URs and the corresponding code content to evaluate the quality of URs against predefined criteria, \ie business context value, completeness, and detail level. After evaluation, the verifier either approves the URs or provides specific improvement suggestions through structured feedback. The Verifier agent can provide feedback to the Search agent and the Writter agent. If the feedback highlights a lack of business knowledge, the Searcher is asked to retrieve additional information, and the Writer revises the URs accordingly. Otherwise, the feedback is sent directly to the Writer for refinement.

Through this iterative collaboration among the three agents, \ours{} can ensure that the generated URs faithfully represent both the technical structure of the project repositories and the broader business needs they aim to fulfill.

\section{Study Design}
To assess the effectiveness of our \ours{}, we perform a study to answer four research questions. In this section, we describe the details of our study, including research questions, studied LLMs, evaluation metrics, evaluated systems, and baselines.  

\subsection{Research Questions and Evaluation Methodology}
Here, we introduce four research questions of this work and propose our evaluation methodology. 

\textbf{RQ1: How does the performance of our \ours{} with different LLMs in generating URs?} This RQ aims to examine the effectiveness of our proposal and its generalization capability on various LLMs in generating URs. In experiments, different LLMs of diverse families are selected for investigation, including GPT-4, Claude 3, and DeepSeek R1. For experimental datasets, we use three existing real-world evaluated systems (Section~\ref{subsec:evalsys}), \ie eTour, eAnci, and SMOS. We use multiple metrics to evaluate the generated URs by \ours{} from the document and set perspective (Section~\ref{subsec:metrics}). As such, we can quantitatively evaluate the performance of \ours{} and compare it with an existing baseline (Section~\ref{subsec:baselines}), \ie HieSum~\cite{dhulshette2025hierarchical}.

\textbf{RQ2: How effectively can our \ours{} against state-of-the-art requirements traceability methods in recovering UR-to-code traceability?} This RQ aims to extensively assess and compare the UR-to-code traceability performance of \ours{} against requirements traceability methods. In this study, five requirements traceability methods are selected for comparison (Section~\ref{subsec:baselines}), including VSM, LSI, COMET, FTLR, and LiSSA. For experimental datasets, we use the same evaluated systems in RQ1, \ie eTour, eAnci, and SMOS. We use multiple widely used metrics (\ie P, R, and F1) to evaluate the quality of the recovered UR-to-code trace links. 

 
\textbf{RQ3: Can \ours{} support end users in validating whether AI-generated repositories align with their intent?} This RQ aims to investigate whether \ours{} can assist end users in assessing the alignment between their intent and the functionalities of AI-generated repositories. To achieve this goal, we conducted a user study involving three practitioners (denoted as A, B, and C). All of them have more than four years of experience in software development and have previously used AI-assisted development tools. In the study, each participant was asked to write a software development intent description. These descriptions were then input into the Kiro tool by the first author to generate corresponding software project repositories. Each participant was then asked to validate the generated repositories without \ours{} and with \ours{}, separately. We recorded multiple metrics (Section~\ref{subsec:metrics}), including validation accuracy, time cost, satisfaction score, and confidence score.


\textbf{RQ4: How efficient is \ours{} in generating URs compared to existing approaches?} This RQ aims to investigate the efficiency of \ours{} by analyzing its runtime and token consumption with the HieSum baseline. To achieve this goal, we measure the total wall-clock time and token usage of our \ours{} across three systems (\ie eTour, eAnci, and SMOS) with three different LLMs (\ie GPT-4, Claude 3, and DeepSeek R1). The result is compared with HieSum under the same hardware and LLM settings. By doing so, we aim to analyze the trade-off between effectiveness and efficiency.

\subsection{Studied LLMs}~\label{subsec:studied_llms}
We introduce three recent LLMs of diverse families. Their detailed information is listed below.

\begin{itemize}
    \item \textbf{GPT-4}~\cite{achiam2023gpt}: GPT-4 is a large-scale transformer-based language model developed by OpenAI. It has demonstrated strong reasoning and generation capabilities across a wide range of natural language processing tasks. GPT-4 is widely regarded as one of the most powerful general-purpose LLMs available and has been applied in domains such as code generation and reasoning. In our study, \textit{gpt-4o-2024-05-13} is used as a representative frontier model of the GPT family to evaluate the performance of our \ours{}.
    \item \textbf{Claude 3}~\cite{claude}: Claude 3 is an advanced LLM released by Anthropic. It emphasizes safety alignment and constitutional AI principles while maintaining competitive performance in reasoning and summarization. It serves as a strong baseline for assessing the generalization capability of \ours{} across different LLM families. In our study, \textit{claude-3-7-sonnet-20250219} is used to evaluate the performance of our \ours{}. 
    \item \textbf{DeepSeek R1}~\cite{guo2025deepseek}: DeepSeek R1 is an open-source LLM that emphasizes reasoning and efficiency. It also demonstrates strong performance in domain-specific reasoning tasks. In our evaluation, \textit{DeepSeek R1} is selected to assess the applicability of \ours{} and to examine whether \ours{} can maintain stable performance with models of different scales.
\end{itemize}

\subsection{Evaluation Metrics}~\label{subsec:metrics}
We use multiple metrics to evaluate the quality of URs and trace links in the above RQs. These metrics are described in the following sections.

\textbf{Metrics for URs from a document perspective.} Following the previous works on code document generation~\cite{yang2025docagent}, we use the LLM-as-a-judge in G-Eval~\cite{liu2023g}. With G-Eval, we consider the following criteria: Completeness (Com): the generated UR document should cover all requirements in the ground truth; Correctness (Corr): the generated URs document should not hallucinate; and Helpfulness (Help): the generated URs document goes beyond merely restating code elements and should elucidate the purpose, usage context, and design rationale.

\textbf{Metrics for URs from a set perspective.} From the set perspective, we treat the generated URs and ground-truth URs as two requirement sets. We then compute Precision, Recall, and F1 to measure the alignment between the two sets. 

\begin{equation}
    \text{Precision} = \frac{|UR_{gen} \cap UR_{gt}|}{|UR_{gen}|}, \quad
\text{Recall} = \frac{|UR_{gen} \cap UR_{gt}|}{|UR_{gt}|}, \quad
F1 = \frac{2 \times \text{Precision} \times \text{Recall}}{\text{Precision} + \text{Recall}} 
\end{equation}

where $UR_{gen}$ denotes the set of generated URs and $UR_{gt}$ denotes the set of ground-truth URs.

\textbf{Metrics for Trace Link.} Unlike traditional requirements traceability, where ground-truth links are defined between fixed requirements and code, our setting involves links between generated requirements and code elements. Since the generated requirements are not identical to the ground-truth requirements, pair-level accuracy computation is infeasible. We therefore adopt a group-based evaluation strategy. Specifically, if a generated requirement links to a set of code elements $C_{gen}$, and there exists a ground-truth requirement that links to a set of code elements $C_{gt}$, we consider the generated link correct if the overlap between the two sets exceeds a given threshold $\theta$.  In our evaluation, we set the threshold as 0.5.

\begin{equation}
\text{Correct}(UR_{gen}) =
\begin{cases}
1 & \text{if } \frac{|C_{gen} \cap C_{gt}|}{|C_{gen} \cup C_{gt}|} \geq \theta \\
0 & \text{otherwise}
\end{cases}    
\end{equation}

From this group-based correctness definition, we compute Precision, Recall, and F1-score for trace link evaluation:

\begin{equation}
    P_{TL} = \frac{|Link{gen}^{correct}|}{|Link_{gen}|}, \quad
    R_{TL} = \frac{|Link{gen}^{correct}|}{|Link_{gt}|}, \quad
    F1_{TL} = \frac{2 \times P_{TL} \times R_{TL}}{P_{TL} + R_{TL}}
\end{equation}

where $Link{gen}^{correct}$ denotes correctly recovered links under the group-based definition, $Link_{gen}$ and $Link_{gt}$ denote the generated trace links and the ground-truth trace links, respectively.  

\textbf{Metrics for User Study.} We evaluate the usability of \ours{} through a combination of objective and subjective measures. On the objective side, we record the validation accuracy (\ie \textbf{Val ACC}), which indicates the proportion of URs correctly validated by participants and the average completion time for each validation task (\ie \textbf{Time Cost}). On the subjective side, we collect participants’ perceptions using Likert-scale questionnaires, focusing on their \textbf{Satisfaction} and \textbf{Confidence} in their validation decisions. These two score range from 1 to 5.


\subsection{Evaluated Systems}~\label{subsec:evalsys}

Our evaluation is based on three real-world software systems: eTour, eAnci, and SMOS. Table~\ref{tab:evalsys} represents a summary of the three evaluated systems. They are provided by the Center of Excellence for Software \& Systems Traceability (CoEST)~\cite{coest} and are commonly used in automated traceability link recovery research~\cite{hey2021improving}~\cite{fuchss2025lissa}. Other datasets for requirements traceability also exist, such as Dronology~\cite{cleland2018dronology}, and LibEST~\cite{libest}. However, we choose these three systems because they provide user-level requirements (\ie use case) as ground truth to address RQ1, and they include trace links between requirements and source code to support RQ2. 

We notice that the natural language are not consistency. The requirements of eTour are written in English, whereas the identifiers in the source code and the names of the use cases are in Italian. In constrast, SMOS contains Italian text for both the use cases and the source code comments, while its identifiers are in English. To ensure linguistic consistency, we follow the treatment in prior studies~\cite{hey2021improving} and translate the identifiers into the predominant language of each project. 


\begin{table}[]
\caption{The datasets used for evaluation, including their domains, artifact types, and statistics.}
\begin{tabular}{ccccccc}
\toprule
\multirow{2}{*}{Dataset} & \multirow{2}{*}{Domain} & \multicolumn{2}{c}{Type of Artifacts} & \multicolumn{3}{c}{Number of Artifacts} \\ \cmidrule{3-7} 
                         &                         & Req                 & Code            & Req        & Code        & Links        \\ \midrule
eTour                    & Tourism                 & Use Case            & Java            & 58         & 116         & 308          \\
eAnci                    & Governance              & Use Case            & Java            & 139        & 55          & 567          \\
 iTrust& Healthcare& Use Case& Java& 131& 226&286\\
SMOS                     & Education               & Use Case            & Java            & 67         & 100         & 1044         \\ \bottomrule
\end{tabular}
\label{tab:evalsys}
\end{table}

\subsection{Baselines}~\label{subsec:baselines}
We select the proposed automated approaches for requirements generation and requirements traceability.

\textbf{Baselines for Requirements Generation in RQ1.} To the best of our knowledge, this is no prior work to explore the user-level requirements generation from the software project repositories. Therefore, we select one similar approaches as baselines. 

\begin{itemize}
    \item \textbf{Hierarchical summarization}~\cite{dhulshette2025hierarchical}: is a two-step hierarchical approach for repository-level source code summarization using individual LLMs. It also prompts the code and business application context for code summarization. Note that although the original approach does not explicitly generate intent-level requirements, it produces package-level requirements descriptions that incorporate business knowledge. In our evaluation, we regard these package-level requirements descriptions as functional descriptions of use cases and compare them against the ground truth.
\end{itemize}

\textbf{Baselines for Trace Links Recovery in RQ2.} In order to evaluate the performance of our \ours{} in recovering trace links between URs and code, we compare it to five state-of-the-art approaches. 

\begin{itemize}
    \item \textbf{Vector Space Model (VSM)~\cite{antoniol2002recovering}}: A classical information retrieval baseline that represents artifacts in a high-dimensional vector space and computes similarity using term frequency and inverse document frequency. Trace links are established based on cosine similarity between requirement and code vectors.
    \item \textbf{Latent Semantic Indexing (LSI)~\cite{marcus2003recovering}}: An extension of VSM that applies singular value decomposition to capture latent semantic structures. By projecting requirements and code into a reduced semantic space, LSI aims to mitigate the vocabulary mismatch problem.
    \item \textbf{COMET}~\cite{moran2020improving}: A Bayesian inference–based method that combines multiple text similarity metrics, integrates developer feedback, and leverages transitive links between intermediate artifacts to refine traceability results.
    \item \textbf{FTLR}~\cite{hey2021improving}: This approach improves traceability by considering finer granularity, mapping requirement sentences to code methods. It employs word embeddings and filtering strategies to enhance the semantic alignment between requirements and code.
    \item \textbf{LiSSA}~\cite{fuchss2025lissa}: A recent retrieval-augmented generation framework that leverages large language models for generic TLR across multiple artifact types. LiSSA combines IR-based retrieval with LLM-based classification and has been shown to outperform traditional IR and heuristic approaches on requirements-to-code tasks.
\end{itemize}


\section{Experimental Results and Analysis}

\subsection{RQ1: How does the performance of our \ours{} with different LLMs in generating URs?}
Table~\ref{tab:ur_generation} demonstrates the experimental results of our \ours{} in generating URs.
We can find that \textbf{(1) \ours{} brings substantial improvements over the hierarchical summarization.} For example, when using Claude 3, the F1 score improves from 0.81 to 0.89 on eTour and from 0.87 to 0.91 on SMOS. Even for GPT-4 and DeepSeek R1, which generally lag behind Claude 3, \ours{} leads to more stable and higher performance across metrics, especially on the SMOS dataset. \textbf{(2) Claude 3 consistently outperforms the other LLMs across all datasets and both approaches.} In particular, under the \ours{} framework, Claude 3 achieves the highest overall scores, with F1 values of 0.89, 0.41, and 0.91 on eTour, eAnci, and SMOS, respectively. These results indicate that Claude 3 is more capable of generating URs that are both correct and helpful, compared to GPT-4 and DeepSeek R1. 

\begin{table}[]
    \centering
    \setlength{\tabcolsep}{3pt}
    \caption{Experimental Results of \ours{} with Different LLMs in Generating URs}
    \begin{tabular}{lcccccccccccc}
\toprule
\multirow{2}{*}{\textbf{Approach}} & \multicolumn{4}{c}{\textbf{eTour}}                                                                                                         & \multicolumn{4}{c}{\textbf{eAnci}}                                                                                                         & \multicolumn{4}{c}{\textbf{SMOS}}                                                                                                          \\ \cmidrule{2-13} 
                                   & \textbf{Com} & \textbf{Corr} & \textbf{Help} & \textbf{F1} & \textbf{Com} & \textbf{Corr} & \textbf{Help} & \textbf{F1} & \textbf{Com} & \textbf{Corr} & \textbf{Help} & \textbf{F1} \\ \midrule
\multicolumn{13}{c}{\textbf{Hierarchical Summarization}} \\ \midrule
GPT-4                              & 2& 1& 2& 0.11& 3& 2& 3& 0.19& 4& 3 & 4 & 0.33\\
Claude 3                           & 4 & 4& 4 & 0.81& 3& 2& 4 & 0.31& 5 & 4 & 4 & 0.87\\
DeepSeek R1                        & 3& 2& 3& 0.21& 3& 2& 4& 0.27& 4& 4& 4& 0.86\\ \midrule
\multicolumn{13}{c}{\textbf{\ours{}}} \\ \midrule
GPT-4                              & 3& 2& 3& 0.23& 3& 2& 2& 0.21& 4& 4& 4& 0.54\\
Claude 3                           & 5& 4& 4& 0.89& 4& 3& 4& 0.41& 5& 5& 4& 0.91\\
DeepSeek R1                        & 3& 3& 3& 0.28& 3& 3& 4& 0.33& 5& 4& 4& 0.89\\ \bottomrule
\end{tabular}
    \label{tab:ur_generation}
\end{table}

\begin{boxK}
\small \faIcon{pencil-alt} \textbf{Answer to RQ1:} \ours{} can bring consistent and substantial improvements for different LLMs of
diverse families on various software systems, showing a powerful generality on various models
and domain-specific systems.
\end{boxK}

\subsection{RQ2: How effectively can our \ours{} against state-of-the-art requirements traceability methods in recovering UR-to-code traceability?}
Table~\ref{tab:trace_links} demonstrates the experimental results of our \ours{} in generating UR-to-code trace links. We can observe that \textbf{(1) \ours{} can improve the quality of trace links than traditional methods.} Across the three datasets, VSM and LSI show very limited performance.  \ours{} consistently achieves significantly higher $F1_{TL}$ values, indicating its ability to capture semantic relevance beyond surface-level textual similarity. \textbf{(2) \ours{} is effectiveness across different LLMs.}  All three LLMs integrated with \ours{} surpass the strongest baselines in at least one dataset. GPT-4 with \ours{} performs best overall, achieving the highest $F1_{TL}$ on eTour and eAnci. Claude 3 with \ours{} also achieves competitive results, while DeepSeek R1 with \ours{} remains slightly behind but still outperforms several baselines. This demonstrates the adaptability of \ours{} to different LLMs while maintaining effectiveness.

\begin{table}[]
    \centering
    \caption{Experimental Results of \ours{} with Different LLMs in Recovering Trace Links}
    \begin{tabular}{cccccccccc}
\toprule
\multirow{2}{*}{\textbf{Approach}} & \multicolumn{3}{c}{\textbf{eTour}}    & \multicolumn{3}{c}{\textbf{eAnci}}    & \multicolumn{3}{c}{\textbf{SMOS}}     \\ \cmidrule{2-10} 
                                   & \textbf{$P_{TL}$} & \textbf{$R_{TL}$} & \textbf{$F1_{TL}$} & \textbf{$P_{TL}$} & \textbf{$R_{TL}$} & \textbf{$F1_{TL}$} & \textbf{$P_{TL}$} & \textbf{$R_{TL}$} & \textbf{$F1_{TL}$} \\ \midrule
VSM                                & .190& .193& .191& .007& .026& .011& .138& .140& .139\\
LSI                                & .207& .211& .209& .007& .026& .011& .103& .105& .104\\
COMET                              & .328& .333& .330& .017& .017& .017& .052& .053& .052\\
FTLR                               & .345& .351& .348& .101& .359& .157& .190& .193& .191\\
LiSSA                              & .381& .219& .278& .155& .158& .157& .121& .123& .122\\ \midrule
GPT 4 + \ours{}                    & .436& .302
& .357& .173& .185
& .179
& .140
& .152
& .146
\\
Claude 3 + \ours{}                 & .403
& .289
& .336
& .149
& .166
& .157
& .128& .131
& .130
\\
DS R1 + \ours{}                    & .369
& .241
& .291
& .158
& .151
& .154& .124& .119
& .121
\\ \bottomrule
\end{tabular}
    \label{tab:trace_links}
\end{table}

\begin{boxK}
\small \faIcon{pencil-alt} \textbf{Answer to RQ2:} \ours{} substantially achieves comparable or superior performance to traditional baselines and recent LLM-based baselines in recovering UR-to-code trace links. In particular, GPT-4 with \ours{} attains the highest overall effectiveness.
\end{boxK}

\subsection{RQ3: Can \ours{} support end users in validating whether AI-generated repositories align with their intent?} ~\label{subsec:rq3}
Table~\ref{tab:user_study} demonstrates the experimental results of our user study. The results show that \ours{} substantially improves the validation process for end users.  In terms of efficiency, the average validation time decreased from 292 seconds to 188 seconds, representing a 36\% reduction. Regarding accuracy, the average validation accuracy increased from 0.63 to 0.85, a gain of 22 percentage points, indicating that \ours{} enables users to make more precise judgments about whether AI-generated repositories align with their intent. Furthermore, subjective feedback also highlights clear improvements: user satisfaction rose from an average of 4.0 to 4.7, while confidence in validation increased from 3.3 to 4.7. 

\begin{table}[]
    \centering
    \setlength{\tabcolsep}{2.5pt}
    \caption{Experimental Results of Our User Study on \ours{}}
    \begin{tabular}{ccccccccccccccccc}
\toprule
\multirow{2}{*}{\textbf{Approach}} & \multicolumn{4}{c}{\textbf{Time Cost (s)}}              & \multicolumn{4}{c}{\textbf{Val ACC}}                & \multicolumn{4}{c}{\textbf{Satisfaction}}           & \multicolumn{4}{c}{\textbf{Confidence}}             \\ \cmidrule{2-17} 
                                   & \textbf{A} & \textbf{B} & \textbf{C} & \textbf{Avg} & \textbf{A} & \textbf{B} & \textbf{C} & \textbf{Avg} & \textbf{A} & \textbf{B} & \textbf{C} & \textbf{Avg} & \textbf{A} & \textbf{B} & \textbf{C} & \textbf{Avg} \\ \midrule
- \ours{}                    & 270
& 310
& 295
& 292
& 0.62
& 0.65& 0.61
& 0.63
& 3
& 3
& 3
& 4.0& 3& 4& 3& 3.3\\
+ \ours{}                       & 180
& 195
& 190
& 188
& 0.85
& 0.88& 0.83
& 0.85& 5& 4& 5& 4.7& 5& 5& 4& 4.7\\ \midrule
Improvement& -90
& -115
& -105
& -104
& -104
& 
+0.23& +0.22& +0.22& +2& +1& +2& +1.7& +2& +1& +1& +1.4\\ \bottomrule
\end{tabular}
    \label{tab:user_study}
\end{table}


\begin{boxK}
\small \faIcon{pencil-alt} \textbf{Answer to RQ3:} t \ours{} effectively supports end users in validating AI-generated repositories. \ours{} reduced validation time by 36\%, improved validation accuracy by 22\%, and significantly enhanced both satisfaction and confidence. 
\end{boxK}

\subsection{RQ4: How much time does \ours{} consume in its execution pipeline?}

Table~\ref{tab:efficiency} presents the execution time and token consumption of \ours{} compared with the hierarchical summarization baseline. Overall, \ours{} introduces additional computational overhead across all three datasets. For example,  the execution time of GPT-4 on eTour increased from 33.98 to 51.12 minutes, and token consumption rose from 197.9K to 327.1K. Similar trends are observed for Claude 3 and DeepSeek R1, where both runtime and token usage are consistently higher. Nevertheless, this increase remains within a practically acceptable range. Even in the most resource-intensive setting (DeepSeek R1 on SMOS), the runtime of 142.73 minutes is still feasible for offline validation scenarios. For LLMs with fast inference speed, the overhead is moderate. For instance, validating eAnci with Claude 3 only took 11.52 minutes. These results suggest that while \ours{} requires more resources, the additional cost does not hinder its applicability and is a reasonable trade-off for the improvements in requirements quality, trace quality, and user experience shown in RQ1-RQ3.

\begin{table}
    \centering
    \caption{Experimental Results of Time and Token Cost}
    \begin{tabular}{ccccccc}
\toprule
\multirow{2}{*}{\textbf{Approach}} & \multicolumn{2}{c}{\textbf{eTour}} & \multicolumn{2}{c}{\textbf{eAnci}} & \multicolumn{2}{c}{\textbf{SMOS}} \\ \cmidrule{2-7} 
                                   & \textbf{Time(Min)}   & \textbf{Token(K)}   & \textbf{Time(Min)}   & \textbf{Token(K)}   & \textbf{Time(Min)}   & \textbf{Token(K)}  \\ \midrule
\multicolumn{7}{c}{\textbf{Hierarchical Summarization}}                                                                                          \\ \midrule
GPT-4                              &                 33.98&                  197.944&                 20.05&                  96.523&                 37.78&                 154.293\\
Claude 3                           &                 14.93&                  208.124&                 7.03&                  99.601&                 12.33&                 159.723\\
DS R1                              &                 134.78&                  302.515&                 55.07&                  144.987&                 86.68&                 236.375\\ \midrule
\multicolumn{7}{c}{\textbf{\ours{}}}                                                                                                     \\ \midrule
GPT-4                              &                 51.12&                  327.138&                 32.48&                  154.276&                 59.90&                 298.354\\
Claude 3                           &                 24.30
&                  361.817&                 11.52&                  165.432&                 21.07&                 276.843\\
DS R1                              &                 175.89&                  518.961&                 91.04&                  285.379&                 142.73&                 434.112\\ \bottomrule
\end{tabular}
\label{tab:efficiency}
\end{table}


\begin{boxK}
\small \faIcon{pencil-alt} \textbf{Answer to RQ4:} \ours{} increases both execution time and token usage compared with hierarchical summarization, with runtime overheads typically in the range of 30–50\% and token costs rising by 60–80\%. However, these increases do not compromise the usability of the approach. 
\end{boxK}




\section{Discussion}

\subsection{Case Study}
\begin{figure}
    \centering
    \includegraphics[width=0.99\linewidth]{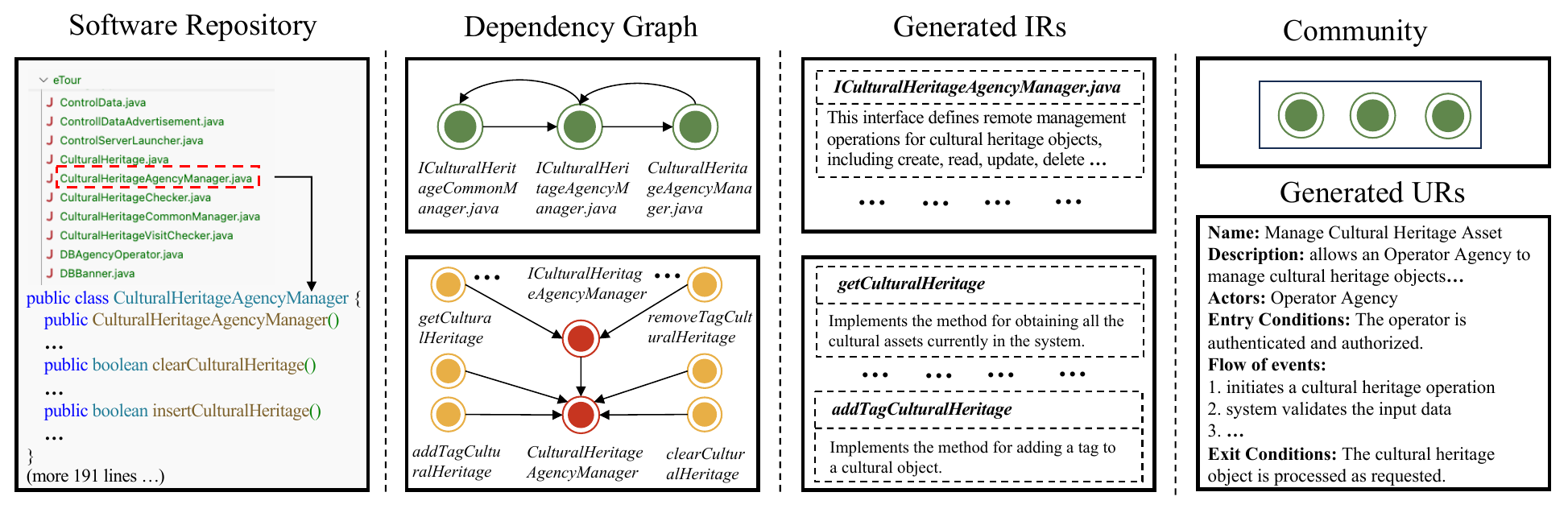}
    \caption{Experimental Results for Our Case Study on the SMOS system.}
    \label{fig:casestudy}
\end{figure}

To evaluate the practicality of our \ours{}, we conducted a case study on the SMOS system. 
As shown in Figure~\ref{fig:casestudy}, our approach transforms the project repository into structured user-level requirements. Starting from the software repository 
(e.g., \texttt{CulturalHeritageAgencyManager.java}), we constructed dependency graphs that capture 
the relationships among code components and files. Based on these graphs, \ours{} generated 
intermediate IRs describing file- and method-level requirements, such as  
\texttt{getCulturalHeritage} and \texttt{addTagCulturalHeritage}. 

We then applied community detection to group related IRs and synthesized them into 
URs. For example, the generated UR (\emph{``Manage Cultural Heritage Asset''}) 
abstracts multiple IRs into a high-level functional goal with actors, entry conditions, flows of events, 
and exit conditions. This case study demonstrates that \ours{} can bridge the gap between 
implementation details and enduser-level requirements, validating its usefulness for 
automated requirements analysis in real-world systems.

\subsection{Threats to Validity}
\textbf{Construct Validity} concerns the relationship between the treatment and the observed outcome. Treat comes from the rationality of the research question we asked. In our study, we aims to design research questions to evaluate the performance of our \ours{}. To achieve this goal, we carefully formulated four research questions targeting different aspects of \ours{}, \ie the quality of generated URs and trace links (RQ1 and RQ2), practical utility in real-world usage scenarios (RQ3), and the efficiency (RQ4). We believe that these questions are well-aligned with our research goals and together provide a comprehensive and balanced evaluation of \ours{}.

\textbf{Internal Validity} concerns the threats to how we perform our study. This validity threat arise from choices in our experimental procedure. The first threat relates to the use of LLM-as-a-judge metrics for evaluating UR documents in RQ1 (Section~\ref{subsec:metrics}). We acknowledge that the design of prompts can affect both the scoring outcome and its reliability. To mitigate this risk, we adopt the prompt template in G-Eval, which has been proven effective in its paper~\cite{liu2023g}. The second threat relates to the replication of VSM and LSI baseline methods used in RQ2 (Section~\ref{subsec:baselines}). The performance of these two methods is sensitive to the threshold used for determining final trace links. To address this, we calculated a project-optimized F1 score by systematically varying the cutoff threshold. This allows us to report the upper boundary of their achievable performance, ensuring a fair and informative comparison. The third threat relates to the potential subjectivity in user study evaluations (Section~\ref{subsec:rq3}). The satisfaction and confidence score are inherently subjective and prone to bias among practitioners. To mitigate this, we invite three experienced practitioners and provide guidelines in advance to standardize their evaluations and reduce personal bias. 


\textbf{External Validity} concerns the threats to generalize our findings. The first threat relates to the selection of LLMs used in our experiments(Section~\ref{subsec:studied_llms}). We select
only three LLMs, \ie GPT-4, Claude 3, and DeepSeek R1. To enhance their representativeness, we ensure diversity in both model family (\ie OpenAI, Anthropic, and DeepSeek) and parameter scale. This design allows us to approximate the performance trends of a broader set of LLMs. The second treat stems from the choice of evaluated systems (Section~\ref{subsec:evalsys}). We conducted experiments on three real-world systems. While these systems are widely used in the literature, there remains a potential threat that the performance of \ours{} may vary on systems with different characteristics or structures. To mitigate this, we selected systems that differ in size, complexity, and domain, providing a reasonable level of diversity.

\subsection{Implications}
We offer practical implications of our \ours{}, highlighting how it can be applied in real-world software engineering workflows and toolchains.

\textbf{Supporting End-User Validation of AI-Generated Software Projects.}
With the rise of LLM-assisted software generation tools, ensuring alignment between user intent and generated project repositories becomes a pressing need. \ours{} empowers non-developer end users to inspect and validate whether the generated repositories truly reflect their original intent by presenting comprehensible URs and their trace links to underlying code components. This reduces reliance on technical intermediaries and enhances transparency in AI-generated software.

\textbf{Facilitating Continuous Maintenance and System Evolution.}
URs and live trace links recovered by \ours{} offer a high-level abstraction layer that remains stable across system changes. This abstraction serves as an anchor for maintainers to understand system functionality, reason about the impacts of code changes, and ensure consistent alignment with business goals. Developers can use these artifacts to support regression testing, impact analysis, and update propagation across requirements, design, and implementation levels.

\textbf{Improving the Explainability of LLM-Based Code Generation Systems.}
By explicitly surfacing the intent-level requirements behind code snippets and mapping them to user goals, \ours{} enhances the explainability of LLM-generated systems. This is aligned with emerging demands for transparency and trust in AI-generated content, particularly in safety-critical and regulated domains such as healthcare, finance, and public governance.

\textbf{Enabling Dataset Construction for RE and Traceability Research.}
\ours{} can be repurposed as a semi-automatic tool to construct or expand datasets for user-level requirements and traceability link recovery research. Researchers can apply it to large repositories to generate candidate URs and trace links, which can then be validated or corrected to form high-quality annotated benchmarks, which addresses the data scarcity issue in requirements engineering research.

\section{Related Work}

\subsection{Code Summarization}
Code summarization~\cite{zhang2022survey} aims to automatically generate concise and human-readable natural language descriptions of code snippets, assisting developers in understanding, maintaining, and documenting software systems~\cite{yau2006some}. These summaries typically describe the functionality of the code and are essential for communicating design decisions and implementation logic~\cite{hu2022practitioners,hu2018deep}. However, as software systems evolve rapidly, manually written comments and documentation often become outdated, inconsistent, or entirely missing. This discrepancy leads to a growing demand for automated summarization techniques.


Early efforts~\cite{gong2022source,hu2018deep,leclair2019neural,lin2021improving} in automated code summarization were inspired by neural machine translation (NMT), where models treated source code as a ``language'' to be translated into a natural language summary. For example, Iyer et al.~\cite{iyer2016summarizing} pioneered the use of LSTM-based encoder–decoder architectures with attention to generate method-level summaries. Building on this foundation, various code representation information is leveraged to further improve NMT-based code summarization performance, such as abstract syntax trees~\cite{gong2022source,leclair2019neural,lin2021improving}, similar code snippets~\cite{li2021editsum,wei2020retrieve}, file context~\cite{haque2020improved}, etc. Furthermore, the pre-training models (\eg Bert~\cite{devlin2019bert}) and fine-tuning paradigm enable fine-tuned summarization tasks with higher accuracy and better generalization. 

More recently, large language models (LLMs) such as ChatGPT~\cite{} and Claude~\cite{} have been applied in zero-shot and few-shot settings. For example, Sun et al.~\cite{sun2023automatic} found that ChatGPT’s zero-shot summaries were generally less precise than those from dedicated models like CodeBERT or CodeT5 with in-distribution summaries. Fried et al.~\cite{fried2022incoder} propose InCoder, a large generative model for code that performs code infilling and synthesis using bidirectional context. They demonstrate the model’s effectiveness through zero-shot evaluation on various code summarization tasks, achieving competitive results. Ahmed et al. ~\cite{ahmed2022few} investigate the effectiveness of few-shot training with LLMs for project-specific code summarization. Interestingly, researchers have also began to explore hierarchical summarization techniques, where summaries are generated at different code granularities (\ie function, file, package). Dhulshette et al.~\cite{dhulshette2025hierarchical} propose a two-step hierarchical approach: first summarize smaller units (\eg functions) using local models, then aggregate those summaries to produce file-level and package-level descriptions. 

Unlike traditional summarization approaches that focus on implementation-level descriptions, our \ours{} targets user-level summaries aligned with the goals of end users. It reframes user-level summaries as a collaborative multi-agent task, involving code analysis, retrieval of domain knowledge, and iterative validation. Rather than producing developer-facing comments, \ours{} outputs user-facing requirement summaries (URs) grounded in context and domain semantics. To the best of our knowledge, this is the first work to explore URs generation from software project repositories.

\subsection{Traceability Recovery}
Traceability recovery~\cite{grundy2002inconsistency} plays a critical role in software engineering by establishing and maintaining connections between different artifacts, such as requirements, design models, test cases, and source code. Given the scale and complexity of modern software systems, manually maintaining these links is often impractical. Therefore, the research community has long sought to automate traceability recovery.

Prior work on automated traceability recovery can be broadly categorized into three categories, \ie information retrieval (IR)-based approaches, machine learning (ML)-based techniques, and LLM-based methods. Early research primarily focused on IR-based methods~\cite{antoniol2002recovering,marcus2003recovering}, which estimate textual similarity between artifacts to infer candidate links. Classical techniques like the Vector Space Model (VSM)~\cite{antoniol2002recovering} and Latent Semantic Indexing (LSI)~\cite{marcus2003recovering} represent artifacts as term-based vectors and compute similarity through statistical co-occurrence patterns. LSI improves upon VSM by applying dimensionality reduction to better capture latent semantic structures. These methods often suffer from vocabulary mismatch issues, where semantically related artifacts share few overlapping terms. To mitigate this, researchers have proposed enhancements such as advanced lexical preprocessing~\cite{cleland2005utilizing,cleland2005utilizing}, integration of dynamic analysis~\cite{dit2013integrating,poshyvanyk2007feature} (\eg execution traces), and incorporating structural dependencies~\cite{kuang2017analyzing} extracted from source code. As data availability and computing power improved, ML-based methods~\cite{guo2017semantically} emerged to learn patterns from labeled trace links. These models treat trace recovery as a classification task, often using handcrafted or learned features from textual and structural attributes of artifacts. Besides, recent work~\cite{fuchss2025lissa} has demonstrated that LLMs augmented with retrieval pipelines outperform traditional IR and ML approaches on linking requirements to code. Despite their potential, LLMs bring new challenges such as hallucination—generating confident yet incorrect links that are not grounded in the actual artifacts. Addressing factual consistency and ensuring trace predictions remain faithful to the source content remains an open research problem.

\ours{} distinguishes itself from prior work in two fundamental ways. First, it is designed to handle scenarios where requirements are missing or incomplete by synthesizing intent-level and user-level requirements directly from the codebase. This allows the system to generate meaningful trace links even in the absence of predefined requirement documents, which is an assumption often implicit in previous approaches. Second, \ours{} supports live and dynamic traceability, meaning that the trace links it generates can evolve alongside the system.

\section{Conclusion}
We proposed UserTrace, a multi-agent system that automatically generates user-level requirements (URs) and recovers live trace links from software project repositories. By coordinating four specialized agents across three phases—repository structuring, IR derivation, and UR abstraction—UserTrace bridges the gap between user intent and code implementation. Experiments on four real-world systems and multiple LLMs show that UserTrace outperforms baselines in both UR quality and traceability recovery. A user study further confirms its effectiveness in helping end users validate whether AI-generated code aligns with their intent. UserTrace contributes to improving software maintainability, explainability, and dataset construction for RE research. Future work includes extending it to more diverse projects and integrating interactive user feedback.

\bibliographystyle{ACM-Reference-Format}
\bibliography{main}

\end{document}